\documentstyle[11pt,paspconf,epsf]{article}
\begin{document}

\title{Cold Gas and Star Formation in Elliptical Galaxies}

\author{Gillian R. Knapp}
\affil{Department of Astrophysical Sciences, Princeton University,
       Princeton, NJ 08544, U.S.A.}

\begin{abstract}
Elliptical galaxies outside dense clusters are observed to contain small
amounts (relative to spiral galaxies) of cold interstellar gas.  This review
discusses the atomic gas, the molecular gas, and the dust in elliptical 
galaxies.  Field elliptical galaxies contain about 0.01 to 0.1 of the
cold interstellar matter content of spiral galaxies of similar luminosity,
and support a low level of star formation.  The surface densities of cold
gas clouds in the centers of some elliptical galaxies are comparable to 
those in the densest regions of the Galactic disk.  These observations
suggest that elliptical galaxies in the field, like spiral galaxies, have a 
long-lived interstellar medium which evolves with the galaxy.  The
timescale for star formation, however, appears to be much shorter, 
leaving these systems with little gas at the present epoch.
\end{abstract}


\section{Introduction}

Cold gas is the `living' part of any galaxy, its bloodstream, so to 
speak.  Stars form from this gas and in turn eject processed material
to the interstellar medium, a symbiotic process which dominates the 
evolution of spiral galaxies to the present day.  Twenty years ago,
elliptical galaxies were thought to be simple systems, consisting
of a single age stellar population formed in a single event at some
long-ago epoch and passively evolving ever since (Tinsley \& Gunn 1976).
The puzzle of what had become of the gas shed by evolving stars (Faber
\& Gallagher 1976) was though to be  solved by the observation of
large amounts ($10^9$ to $10^{10} ~ M_{\odot}$) of hot ($10^6$ to $10^7$
K) gas in extended halos around elliptical galaxies by the EINSTEIN
and, more recently, ROSAT and ASCA observatories (e.g. Canizares
et al. 1987). However, as shown by several papers at this conference, 
for example, the
observed gas accounts for only a small fraction of the expected total. 
Hot gas does not directly support
star formation.  Nevertheless, as the discussion at this conference
shows, there is not yet a comfortable consensus on the origin of
the hot gas, nor on the fate of the gas produced by stellar evolution,
nor on the relationship between cold and hot gas in these systems.

Even in 1976, it was known that some elliptical galaxies contain
substantial amounts of cold interstellar gas, visible as dust lanes
and patches: the paradigm galaxy is NGC 5128, the powerful nearby
radio galaxy Centaurus A.  However, since the morphological
definition of elliptical galaxies includes ``smooth appearance''
with no discernible structure, such galaxies were classified as 
``Ep'', ``SO'', ``SOp'', or ``I0'', where p = peculiar.  Galaxies
like NGC 5128 were considered to be involved in a merger with a
gas-rich spiral galaxy and to be highly atypical elliptical galaxies,
or not elliptical galaxies at all.  

A lot has changed in the last 20 years, as the observational data at
all wavelengths have improved dramatically.  Not only has it
been found that the stellar populations and abundances in elliptical
galaxies trace a varied and complex star formation history, but
tracers of ongoing star formation and of many phases of the 
interstellar medium have been found in a large enough number of
systems that these might be regarded as part of the `typical'
elliptical galaxy.  Nevertheless, both current star formation
activity and cold interstellar gas are present in much smaller
amounts in elliptical galaxies and hence are much more difficult
to distinguish against the galaxy's background - paradoxically, 
these phenomena are easier to measure in spiral galaxies because they
dominate the radiation from the typical spiral galaxy at essentially 
all wavelengths.

H$\alpha$ emission from warm photoionized gas is a reliable tracer of star
formation activity in spiral galaxies, but in ellipticals is also produced
by powerful AGNs and by photoionization by the old stars.  Hot ($\geq
10^6$ K) gas is produced by supernova shock heating in spiral galaxies and 
coexists intimately with the cold ISM - but any such gas in elliptical
galaxies is completely insignificant compared to the hot 
gas halos.  Diffuse weak radio continuum emission reliably
traces star formation in the disks of spiral galaxies, but is insignificant 
compared to the powerful radio emission from AGNs, which are present in almost
all elliptical galaxies.  And finally, the spectroscopic and photometric
signatures of recent star formation are hard to distinguish against
the bright ambient stellar population.

Cold gas itself is more straightforward.  HI, usually in very extended
structures, has been mapped via its 21 cm emission from several tens
of elliptical galaxies.  Molecular gas is detected in a surprisingly
large fraction of elliptical galaxies via its emission in the CO 
rotational lines (and, in a few galaxies, via several other molecular
probes).  Dust is detected in lanes in optical color images, via polarization
and color gradients
in the optical light (Wise \& Silva 1996),
and via emission at mid- and far-infrared 
and submillimeter wavelengths.  The most sensitive probe is probably
fine structure
emission from [CII] and [OI] (Malhotra et al. 1998).  Together, these
observations show that elliptical galaxies, like spiral galaxies, contain
an active cold interstellar medium.  This review will discuss these probes of 
the interstellar medium in turn; the amount, distribution and origin
of the interstellar gas, and its relationship to the evolution of elliptical
galaxies.  It will not discuss the use of HI maps to measure the mass
distributions of elliptical galaxies, and
will mention only in passing cold ISM at high redshifts and the ISM in radio
galaxies, two areas in which a lot of important and vital work has been done
recently.

\section{HI in Elliptical Galaxies}

This subject has really taken off in recent years, thanks to deep
single-dish surveys (Huchtmeier 1994, Huchtmeier et al. 1995), and large scale
mapping observations using the large southern and northern radio synthesis 
arrays, the Australia Telescope (AT) and the Very Large Array (VLA). 
The exciting work from the former telescope is described at this
conference in papers by T. Oosterloo and R. Morganti.  The general subject, 
as well as recent VLA work, is discussed in recent reviews by Schiminovich
et al.  (1998a,b),  van Gorkom (1998), and van Gorkom \&
Schiminovich (1998).

\begin{table}
\caption{HI Detection Rates in Early-Type Galaxies.} \label{tbl-1}
\begin{center}
\begin{tabular}{crrrrrrrrrrr}
Type & Number & Detected \\
\tableline
E & 64 & 5\% \\
E/S0 & 23 & 17\% \\
S0 & 103 & 20\%  \\
Ep, S0p & 20 & 45\% \\
S0a, S0ap & 35 &  43\% \\ 
Sa, Sap & 103& 78\% \\
\end{tabular}
\end{center}
\end{table}

Table 1 shows the detection rates for early-type galaxies from the 
Revised Shapley-Ames Catalogue (Sandage \& Tammann 1981) compiled by
van Gorkom (1998).  The detection rate rises steadily towards later-type
galaxies, and is very small for elliptical galaxies.  The detection rate is 
far higher for `peculiar' galaxies, which are strongly preferentially
to be found in the field.  Mapping observations with the AT of
dwarf and dust lane elliptical galaxies (Oosterloo, this conference: Morganti,
this conference) and of shell galaxies by Schiminovich et al. (1994, 1995,
1998a,b,c) give high detection rates - 8/11 in the former case, 12/22
in the latter, statistics from van Gorkom (1998).  Early synthesis array
HI maps of HI rich elliptical galaxies often showed warped and inclined
structures, suggestive of recent acquisition; however, the high HI
detection rate, and the regularity of HI disks and rings in many 
other galaxies, is
more compatible with an intrinsic origin for the HI.  It will be
argued in subsequent sections that the dust and molecular gas content
of elliptical galaxies also support this contention: but the particular
peculiarity of HI in elliptical galaxies is its large observable extent.
Many pieces of evidence for spiral galaxies suggest continued infall of
gas from large radii - this infall may also be part of the evolution of 
elliptical galaxies and responsible for the large HI structures.  Where
would such gas come from?  Dense clusters of galaxies today contain 5 - 10
times as much mass in hot intercluster gas as in stars, with the total
amount of normal matter more or less in agreement with the amount at
high redshift measured by Ly-$\alpha$ absorption.  However, in
the field, only a fraction (1/5 or so) of that mass is presently in 
stars.  Cen \& Ostriker (1998) suggest that most of the normal matter
in the field in the present-day Universe is in the form of diffuse gas heated
to temperatures in the hard-to-observe range of $\rm 10^5 ~ - ~ 10^7$ K by 
gravity.  If this reservoir provides gas infalling to spiral
galaxies, then such gas should also be accreted by elliptical galaxies
in the field, and may be the source for the large HI structures and be part
of a continually-evolving interstellar medium.

\section{Dust:  Far-Infrared and Submillimeter Emission}

Interstellar dust heated by starlight radiates at far infrared wavelengths,
and dominates the emission from essentially all spiral galaxies at
wavelengths longer than a few $\rm \mu m$.  Long wavelength
emission is also detected towards a significant fraction of
early-type galaxies by IRAS (12$\rm \mu m$, 25$\rm \mu m$, 60$\rm
\mu m$ and 100$\rm \mu m$), ISO (between about 5$\rm \mu m$ and
200 $\rm \mu m$; Malhotra et al. 1998,  S. Madden, this conference) and at 
submillimeter wavelengths.  The largest sample available is from the IRAS
survey.  These data suggest the presence of small amounts of interstellar
dust, and hence of cold interstellar gas, in many nearby 
elliptical galaxies.  However, the data are difficult to interpret
because the observed flux densities are low and at only a few times
the r.m.s. noise level, and are difficult to
distinguish from other sources of emission.  At 12$\rm \mu m$ and 25$\rm
\mu m$, emission can arise from hot (small) interstellar dust grains;
hot (several 100 K) dust in the vicinity of an AGN (these being 
common in elliptical galaxies); stellar photospheres; and circumstellar
dust produced by mass loss. In most elliptical galaxies, the bulk of the
emission at these wavelengths seems to be due to stars and circumstellar dust
(Knapp et al. 1992).
At 60$\rm \mu m$ and 100$\rm \mu m$, emission 
can arise from cool interstellar dust, non-thermal synchrotron emission from
an AGN (cf. the discussion of M 87 by Knapp et al. 1990 and Braine \& Wiklind
1993), confusion with nearby dust - rich galaxies, and structure in the 
Galactic cirrus.  Knapp et al. (1989) report the detection of 60$\rm
\mu m$ and 100$\rm \mu m$ emission
from some 40\% of elliptical galaxies brighter
than $\rm 14^m$, in reasonable agreement with the fraction of elliptical
galaxies in which optical dust lanes are seen (Ebneter at al. 1988;
van Dokkum \& Franx 1995); 
the infrared
fluxes show no correlation with optical brightness.
However, because of the several possible/likely sources of contaminating
emission, this is an upper limit on the percentage of galaxies with
detectable dust emission - Bregman et al. (1998) find that the
percentage is much lower,  about 15\%.

Broad-band spectral energy distributions have been measured for several
galaxies using IRAS, ISO and the submillimeter telescopes (IRAM, JCMT
and CSO) (e.g. Wiklind \& Henkel 1995; Knapp et al. 1998a). 
These observations show that the
emission, including the submillimeter emission, comes from the inner
regions ($\rm \leq r_e$)of the galaxies, and that there is no evidence
for cold dust.  Cold dust at large distances from the center of the
galaxies is by no means ruled out, but the submillimeter emission which
is {\it observed} comes from the inner regions.  The inferred column densities
of cold ISM are high, $\rm > 10 ~ M_{\odot} ~ pc^{-2}$ in some cases
(cf. the CO observations of NGC 3928 by Li et al. 1994).
This is higher than in the Galaxy's molecular
ring. The small central disks recently found in the inner regions of early type
galaxies also have very high central surface densities (e.g. Scorza \&
van den Bosch 1998), and may have formed from these dense gas structures.

A rather odd correspondence between HI and infrared emission is found.
Galaxies with detectable 60/100$\rm \mu m$ emission are much more likely to
have detectable HI, and vice-versa; but the fluxes are essentially
uncorrelated.  As we discuss below, the dust and molecular gas are
quite tightly correlated.  A likely explanation is that
the presence of any one of these indicators:
dust absorption or emission, HI, or CO, in a galaxy shows that 
cold interstellar gas is present in all its phases;
but since the distributions of the atomic
and molecular gas are different, the total amounts observed are not well 
correlated.

\section{Molecular Gas}

The subject of molecular gas in elliptical galaxies has been reviewed
in detail by Henkel \& Wiklind (1997) and by Rupen (1998).  As is the
case for spiral galaxies, the workhorse molecule is CO.  Several largish
surveys of nearby early-type galaxies in the CO(1--0) or CO(2--1) lines have
been completed in recent years by Wiklind \& Henkel (1989), Lees et al. (1991),
Wiklind, et al. (1995), and Knapp \& Rupen (1996), and CO
emission has been detected from some tens of galaxies.  In addition, possible
CO {\it absorption} has been seen in a few galaxies with flat-spectrum
compact radio nuclei, most notably NGC 5128 (Israel et al. 1991).  The
CO(2--1) observations of NGC 1052 reported by Wiklind et al.
(1995) and Knapp \& Rupen (1996) tentatively show CO
absorption towards the nucleus, and the presence of dense
molecular gas in its inner regions is also shown by the detection and
mapping of powerful $\rm H_2O$ maser emission (Claussen et al. 1998).

It should be emphasized that, like the infrared emission, the CO data 
for elliptical galaxies are not,
at present, individually very reliable (with a few exceptions).  The lines
are weak, and are typically observed with spectrometers covering only
about 700 $\rm km~s^{-1}$, so that any very broad line emission is
likely not to be detected.  Some detections have been confirmed using 
several different telescopes, and some few galaxies (see, e.g. the
papers by L. Young and J. Cepa, this conference) have been mapped
in detail with interferometers.  However, many other detections are
tentative; indeed, the field awaits the construction of sensitive and
versatile arrays such as the Millimeter Array, which is to be built on
Cerro Chajnator, Chile.  Nevertheless, the presence of dense molecular
gas in a significant number of elliptical galaxies is secure.  Table 2
shows the detection statistics, from the concatenation of the above surveys
and other CO observations of early-type galaxies, compiled 
by Rupen (1998).  Column 1 contains
the galaxy morphological type, column 2 the number of galaxies
observed, and column 3 the percentage detection rate.  In keeping with 
the discussion in Section 3, the ``peculiar'' Es and S0s are folded 
in with other members of the class.  Table 2 shows that the detection
rates rise steadily towards later morphological types.  If the
strength of the CO emission is related to molecular gas in elliptical
galaxies in the same ratio as is inferred for spiral galaxies, the
molecular gas masses range from $\rm 3 \times 10^5 ~ M_{\odot}$
to several $\rm \times 10^8 ~ M_{\odot}$.  Lees et al. (1991)
show that the molecular gas mass normalized to the blue luminosity
also increases steadily towards later morphological types. 

\begin{table}
\caption{CO Detection Rates in Early-Type Galaxies.} \label{tbl-2}
\begin{center}
\begin{tabular}{crrrrrrrrrrr}
Type & Number & Detected \\
\tableline
E & 61 & 39\% \\
dE & 4 & 100\% \\
E/S0 & 26 & 31\% \\
S0 & 43 & 47\% \\
S0a & 22 &  55\%\\ 
\end{tabular}
\end{center}
\end{table}

\begin{figure}
\plotone{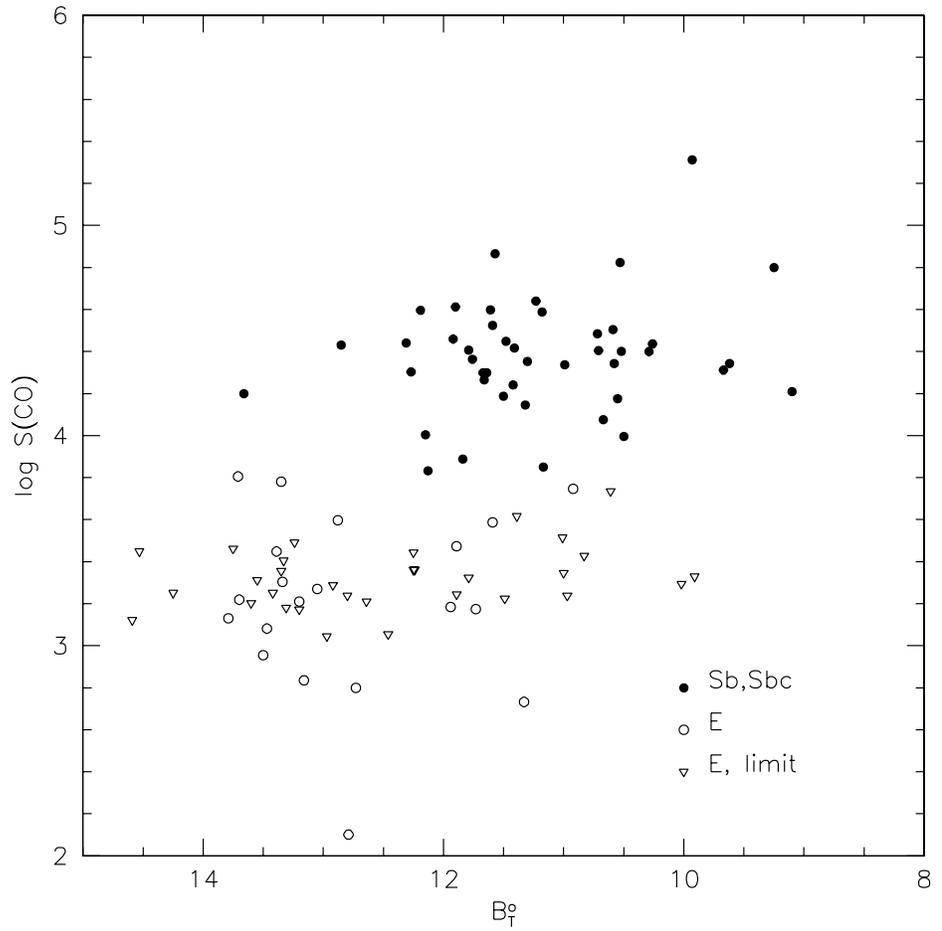}
\caption{CO line flux (in units of $\rm K \times km~s^{-1} ~ arcsec^2$,
see text) versus total corrected blue magnitude $\rm B_T^o$ for samples
of spiral and elliptical galaxies.  The filled circles show the data
for the Sb/Sbc galaxies from the sample observed by Elfhag et al. (1996).
The open symbols show the data from the sample of elliptical galaxies
observed by Lees et al. (1991) and Knapp \& Rupen (1996).  The inverted
triangles show the upper limits on the CO flux for galaxies in which CO
was not detected: the limits on the line intensity are calculated as
the channel-to-channel r.m.s. noise multiplied by 300 $\rm km~s^{-1}$.}
\label{fig-1}
\end{figure}

Figure 1 shows the comparison between the molecular gas content of 
elliptical and spiral galaxies; the CO line flux S(CO) is
plotted against the total blue magnitude $\rm B_T^o$.  The CO line flux is 
in rather arbitrary units: S(CO) = $\rm I(CO) \theta^2$, where I(CO)
is the integrated line intensity in $\rm K \times km~s^{-1}$ and $\theta$
is the half-power beamwidth of the telescope used to make the observations.
The comparison sample of spiral galaxies consists of the Sb/Sbc galaxies
from the survey of Elfhag et al. (1996), while the CO data are from 
observations with the Caltech Submillimeter Observatory by Lees et al. 
(1991) and Knapp \& Rupen (1996).  The CSO observations are made in the
CO(2--1) line with a half-power beamwidth of $\rm 30''$, and were
made only at the central position of the galaxy.  The Elfhag et al.
observations were made with beamwidths of $\rm 33''$ or $\rm 44''$
of the CO(1--0) line, and again were made only at the central
position of the galaxy.  The use of the CO(1--0) line for one sample and
the CO(2--1) line for the other should not pose too much of a problem,
because in Galactic molecular clouds, at least, the brightness temperatures of
molecular clouds in the two lines are essentially equal. 

Figure 1 shows that, on the average, the CO fluxes
are 20 - 30 times lower in elliptical galaxies than in spiral galaxies
of the same total blue magnitude - elliptical galaxies contain a 
lot less dense cold gas.  Note also the large scatter in the CO flux
densities of the elliptical galaxies and the complete lack of correlation
between the CO flux and the blue light.  Many of the CO observations
have resulted only in upper limits on the CO fluxes, as shown in Figure
1.  The values of the upper limits overlap completely with those of the
detections; statistical analysis of data sets with this sort of mix of
detections and upper limits (see Lees et al. 1991, for example) 
show that the upper limits are not restrictive: in other words, the data
are consistent with all elliptical galaxies having weak CO line
emission and hence small amounts of molecular gas.

Figure 2, where the CO fluxes are plotted versus the 100$\rm \mu m$
fluxes for the same two samples of galaxies, tells a different 
story. Although the beam used to observe the 100$\rm \mu m$ flux densities
is large (about $\rm 3' \times 7'$) and therefore observes most of the 
100$\rm \mu m$ emission (and further the 100$\rm \mu m$ data plotted in
Figure 2 include total flux densities for extended galaxies, taken from
NED), the CO and 100$\rm \mu m$ emission are fairly tightly correlated
for the spiral galaxies, unlike the case with blue light (Figure 1).
Figure 2 suggests a fairly well-mixed 
dense interstellar medium in these galaxies,
with a roughly constant gas to dust ratio.  The same data are also plotted
for the sample of elliptical galaxies.  Although the scatter is much larger 
for these low signal-to-noise ratio observations, the measurements for
the elliptical galaxies lie on the same relationship as
found for spiral galaxies.  The dotted line in Figure 2 shows a relationship
of slope 1, i.e. $\rm S(CO) ~ \propto ~ S(100 \mu m)$.  Thus most of
the 100$\rm \mu m$ flux in these systems is associated with the 
molecular interstellar medium, and the global characteristics of the
dense ISM are roughly the same in elliptical galaxies as they are in 
spiral galaxies.

\begin{figure}
\plotone{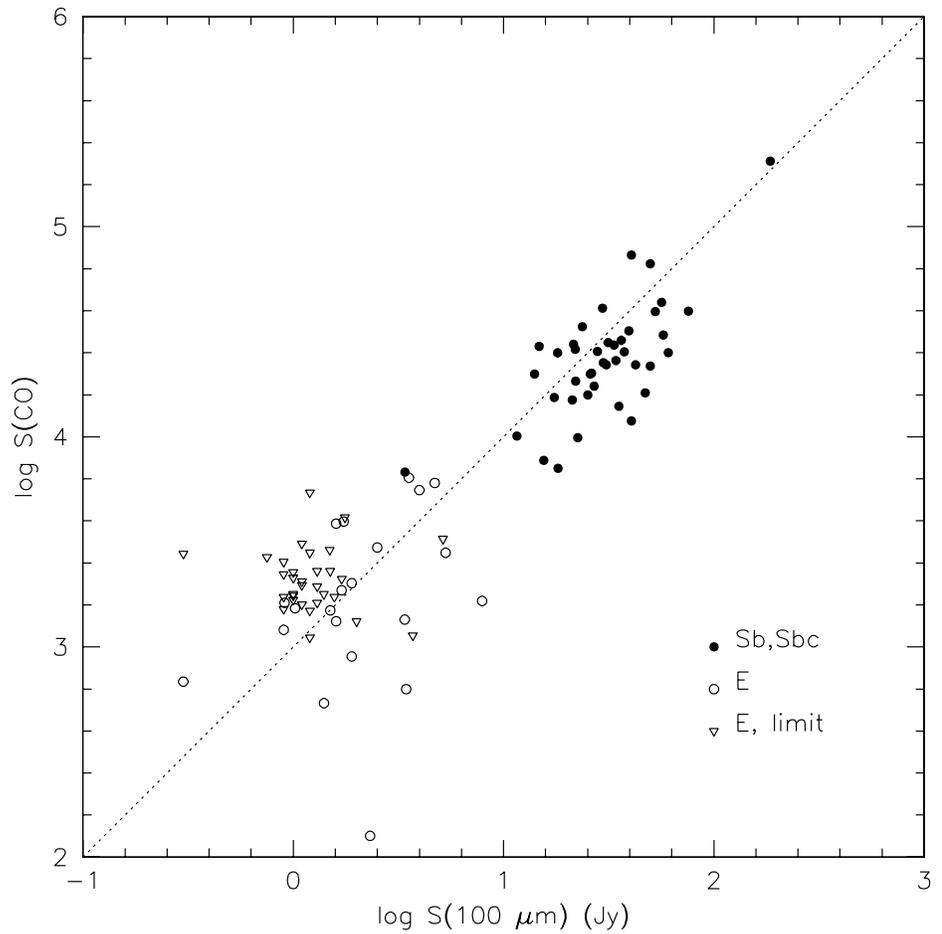}
\caption{CO line fluxes for spiral and elliptical galaxies versus 100
$\rm \mu m$ flux densities from IRAS.  The symbols are as for Figure 1.}
\label{fig-2}
\end{figure}

\section{The Relationship Between Hot and Cold Gas}

Figure 3 shows $\rm L_X/L_B$ versus $\rm L_{100}/L_B$ for nearby
elliptical galaxies.  The X-ray emission is assumed to come from 
hot gas and the 100$\rm \mu m$ emission from cold dust (although
see below).  The X-ray luminosities are from the EINSTEIN X-ray data 
compilation of Fabbiano et al. (1992, FKT), and the 
100$\rm \mu m$ data from Knapp et al. (1989).  The data plotted in Figure 
3 are for galaxies of morphological type E or E/S0 for which data are given in
both catalogues and which are detected at either or both X-ray and
infrared wavelengths.  

The X-ray emission can arise from the stellar component, from AGNs, and
from hot gas.  The stellar component was approximated by scaling X-ray to blue 
light flux for spiral galaxies, also using the data in FKT: $\rm
L_X^*(erg ~ s^{-1}) ~ = ~ 1.5 \times 10^{30} ~ L_B/L_{\odot}$
(cf. Brown \& Bregman 1998) (this includes a factor of 3 for the mass
to light difference between the stellar populations of elliptical 
and spiral galaxies).  If the `stellar component' is greater than
or equal to the observed X-ray flux, it is taken as an upper limit to
the X-ray flux - hence the horizontal line of X-ray upper limits
in Figure 3.  The infrared luminosities (in $\rm erg ~ s^{-1}$)
are calculated from $\rm L_{100} ~ = ~ \nu S_{\nu} \times 4 \pi D^2$
using the distances given by FKT.

Figure 3 shown no proportionality between the X-ray and infrared
luminosities.  The galaxies detected at both X-ray and infrared
wavelengths show, if anything, anti-correlation between the fluxes.  Two
galaxies are obvious exceptions to this, with X-ray detections
and $\rm L_{100}/L_B ~ \sim$ 32.15 (Figure 3).  These galaxies,
NGC 3258 and NGC 3894, have AGNs, and it is likely that AGN emission
at significant or dominant levels is also present in some other galaxies
plotted in Figure 3.  

With the exception of the above two galaxies, those with relatively large
values of $\rm L_{100}/L_B$ are not detected at X-ray wavelengths, while those
with low values are.  If the sample is divided at its median 100$\rm
\mu m$ ratio, $\rm L_{100}/L_B$ = 31.4, the X-ray detection rate is 
80$\pm$20\% for galaxies with low infrared flux and 30$\pm$20\% for 
galaxies with high infrared flux.  Despite probable contamination
from AGNs, these data show that cold and hot gas are anti-correlated.

A similar result is given by an exhaustive study of the interstellar
gas and radio emission from both elliptical and S0 galaxies by 
Eskridge et al. (1995a,b).  Braine \& Wiklind (1993) and Braine et
al. (1997) find no CO emission to very low limits for X-ray bright
elliptical galaxies, both in and out of clusters, in contrast to
the CO detection rate in field elliptical galaxies discussed above.
Maps of the cold and hot ISM in the bulge-dominated Sa galaxy NGC 1291
by Bregman et al. (1995) find that the cold gas is associated with
the disk while the hot gas is associated with the bulge, while
an analysis of the hot and cold gas contents of E, S0 and Sa
galaxies by Roberts et al. (1991), Bregman et al. (1992) and Hogg
et al. (1993) finds the same result.  The 
anti-correlation between hot and cold gas found for elliptical 
galaxies (which have no disks),
however, suggests a more
intimate relationship between the hot and cold gas in these 
systems; the destruction of cold gas
by hot interstellar gas, or the cooling of hot gas if enough
cold gas is already present.  

\begin{figure}
\plotone{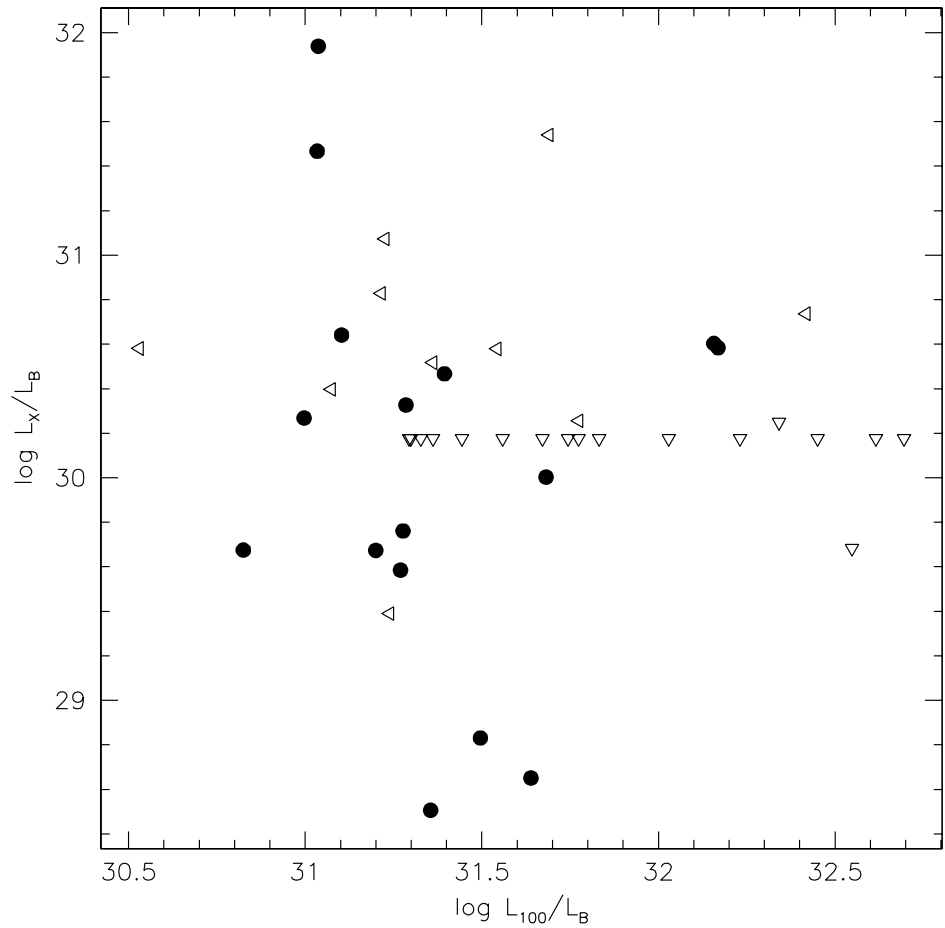}
\caption{X-ray versus infrared emission from elliptical galaxies (see
text).  Filled symbols: galaxies detected at both X-ray and infrared
wavelengths. Open triangles: 3$\sigma$ upper limits for either X-ray
or 100$\rm \mu m$ emission.} \label{fig-3}
\end{figure}

\section{The Distribution of Dust in Elliptical Galaxies: Comparison of
Far Infrared Emission and Optical Extinction}

Dust is observed in optical images of galaxies because it absorbs and
reddens the starlight.  The presence 
of dust lanes and clouds in
some elliptical galaxies has long been known (Hawarden et al. 1981; 
Ebneter et al. 1988), and galaxies with prominent dust lanes tend to be
HI rich (Morganti, this conference). Systematic optical searches for dust
in elliptical galaxies (Goudfrooij \& de Jong 1995; van Dokkum \& Franx
1996, DF) find dust lanes in the inner regions of a sizable fraction
(about 50\%) of elliptical galaxies.  The analysis of HST archive
data by DF is particularly compelling because these elliptical galaxies
were selected for observation to be as far as possible dust-free, since
the object of the observations was the study of nuclear structure.

The two methods for finding dust clouds and measuring their mass,
optical extinction and far-infrared emission, have both advantages
and disadvantages. Optical measurements can find small amounts 
($\rm \leq 1000 ~ M_{\odot}$) of dust and can measure reddening
and hence dust properties, but they are sensitive to geometry: edge-on
disks are easy to see, face on disks are not; and the dust cannot be
traced much beyond the bright inner (within about $\rm r_e$) regions.
Far infrared measurements are unaffected by geometry but have three
disadvantages; they are far less sensitive than are optical observations;
and since the emission is fairly weak, there are several sources of confusion,
as discussed in Section 3, so that the measured dust masses are
frequently unreliable.  Knapp et al. (1998b) discuss
and compare the optical and infrared observations of the DF sample
of elliptical galaxies, and find (1) dust is directly detected in about
80\% of the galaxies, in good agreement with the DF detection rate
corrected for projection effects; (2) the dust masses inferred from
the far infrared measurements are almost always much higher than
those measured by optical observations (cf. Figure 4; similar results
are found by Goudfrooij et al. 1994 and by Bregman et al. 1998); and
that the {\it surface densities} of dust measured at optical and infrared
wavelengths are in remarkably good agreement (Figure 4). (In these figures,
NGC 2110 is a galaxy from DF's sample which is not classified as an 
elliptical galaxy in any of the conventional catalogues, and is far dustier
that any of the other galaxies in the sample).

\begin{figure}
\plottwo{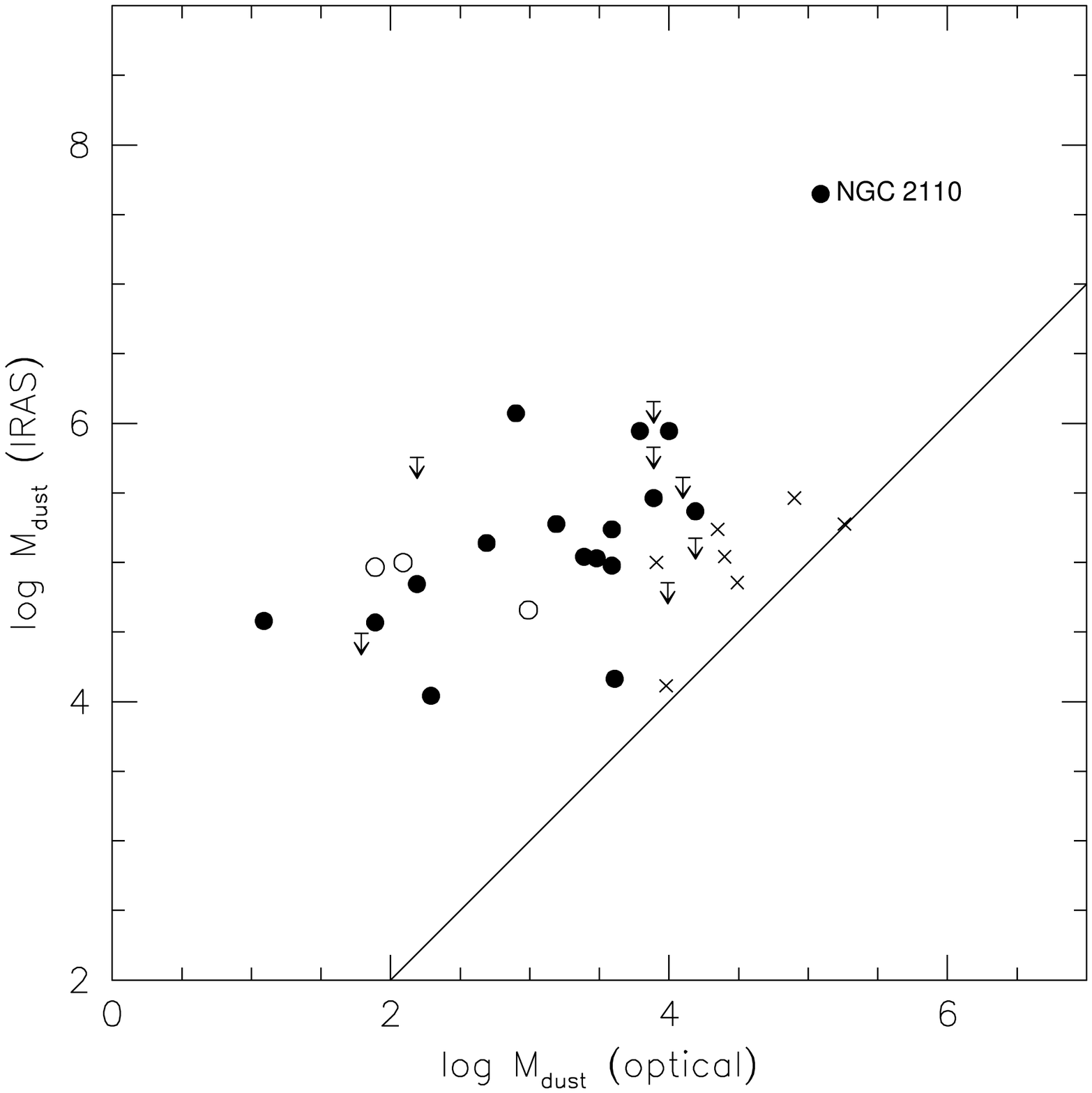}{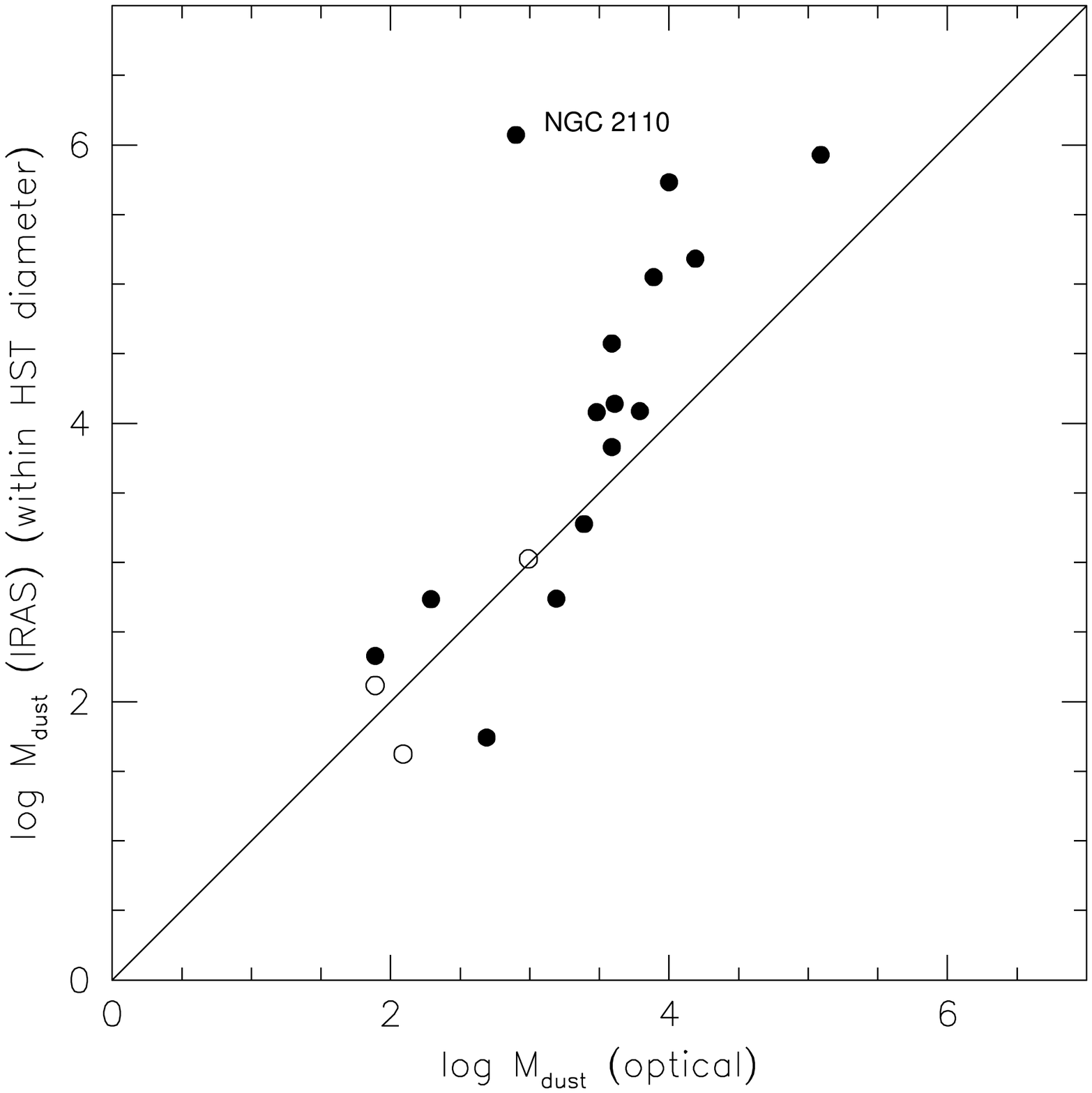}
\caption{Comparison of optical and infrared measurements of the dust 
mass in a sample of HST observed elliptical galaxies (see text).  Left
panel: total mass.  Right panel: mass within the optically-
determined dust radius. The diagonal lines denote equal dust masses.
Open symbols: galaxies detected at only one IRAS wavelength. Crosses:
galaxies with ground-based measurements by Goudfrooij et al. (1994).}
\label{fig-4}
\end{figure}

Thus the `typical' elliptical galaxy appears to contain a small amount of
interstellar dust, and by inference cold gas,
but it should be noted that neither the optical nor the infrared
observations are at all sensitive to dust at large distances from the
center of the galaxy, and that many of these galaxies could contain far more
interstellar dust than is seen by these observations.  The question
of whether the extended HI disks or rings seen around many elliptical
galaxies contain dust or are primordial thus remains open.

The high resolution optical data discussed by DF were originally acquired to
investigate core properties (Lauer et al. 1995).  These investigations show
that elliptical galaxies fall into two classes - the more luminous have
boxy isophotes and cores, while the lower luminosity galaxies have
disky isophotes and no cores.  Interestingly, the incidence of
detectable interstellar dust is essentially identical in these two
types of galaxy.

\section{Star Formation}

Star formation is seen directly in the nearby dusty elliptical 
galaxy NGC 5128.
A small number of low luminosity elliptical galaxies (e.g. NGC 855, 
NGC 2328, NGC 3928, and NGC 5666) show many of the signs of star formation
or even a starburst: blue colors, relatively high dust temperature, in
some cases weak extended non-thermal radio emission, and HII-region like
emission spectra.  In terms of these tracers, the star formation seems to
be behaving just as it does in the disks of spiral galaxies.  The 
remarkable thing that elliptical galaxies have to contribute is the 
tiny amount of gas that seems to be required to support full-blown
star formation.  NGC 855, for example, has $\rm 4 \times 10^7 ~ M_{\odot}$
of HI and only $\rm \sim 10^6 ~ M_{\odot}$ of $\rm H_2$, the amount found
in one Galactic Giant Molecular Cloud; yet it is cheerfully forming
stars, as shown by its blue color and emission line spectrum.  Young
(this conference) finds that the Local Group ellipticals NGC 205 and
NGC 185 share this characteristic.  Thus star formation seems to be 
essentially a local process - all it takes is a few thousand $\rm M_{\odot}$
of molecular gas.

These star formation tracers, if they are present at all, are 
swamped by other sources
of emission in most bright elliptical galaxies, as discussed in Section 1.  It
is nevertheless highly likely that many other elliptical galaxies support
star formation at a low rate - the densities and column densities
inferred from CO and infrared observations are high in many cases.

A set of beautiful ISO LWS spectroscopic observations by Malhotra
et al. (1998) finds [CII] and [OI] emission from several elliptical
galaxies, showing the presence of at least several $\rm \times 10^5
~ M_{\odot}$ of gas in these galaxies.  While these lines, especially the
[CII] 158 $\rm \mu m$ line, are thought to be important or dominant
coolants of the diffuse ISM, ISO and airborne observations of 
star forming regions show that much of the emission also arises from
photo-dissociation regions, and the detection of these lines in
elliptical galaxies probably demonstrates the presence of star formation at a
low level.  Alas, ISO is no more and there is no long-wavelength
spectroscopic capability on SIRTF, so it will be a long time before
these lines can be probed again in early-type galaxies.

\section{The Evolution of the ISM in Elliptical Galaxies}

The observations discussed in the previous sections strongly suggest 
that many elliptical galaxies have a small amount of cold interstellar gas,
whose global properties (admixture of gas and dust, molecular and 
atomic medium, etc.) are not dissimilar to the ISM in spiral galaxies -
there is just a lot less of it.  This rather strongly suggests that 
the `origin' of the cold gas in elliptical galaxies is the same as that
of spiral galaxies - to wit, the present-day gas content and its
composition is the result of continuous evolution: initial
star formation is incomplete, stars continue to form, gentle and 
violent mass loss from stellar evolution enriches the gas and
injects dust, and the ISM is partly replenished by mass loss, the 
infall of (perhaps) primordial gas and gas-containing companions.
The ISM, in other words, does not have a single origin any more than
that in a spiral galaxy does - it has evolved along with the rest
of the galaxy.  This point of view suggests that elliptical galaxies
have always contained a cold ISM, and that its presence is not an anomaly. 

Among the more exciting developments in recent years is the plethora
of data on the properties of very distant galaxies.  Many of these
observations have shown that many elliptical galaxies
had much more interstellar gas and star formation in the past than they
do now.  An investigation of the colors of gravitationally lensed quasars by
Malhotra et al. (1997) shows that the lensed objects 
identified in radio surveys are systematically much redder, by 
several magnitudes in some cases, than are lensed objects identified
in optical surveys, suggesting the presence of several magnitudes
of extinction in a large fraction of lenses.  Most of the lenses
are considered to be elliptical galaxies, as these systems have the
mass and central density necessary to produce observable splittings.  Thus the
typical elliptical galaxy at z $\sim$ 0.5 often has a large amount
of dust in its inner regions. 
A range of epochs for star formation may be inferred from color magnitude
diagrams and other studies which find differences in the stellar
populations among galaxies - for example, the recent study by Pahre
et al. (1998) which shows that the slope of the elliptical galaxy 
sequence (the fundamental plane) depends on the photometric band
in which the measurement is made, illustrating the difference in
stellar content along the sequence.

How are these results related to the present-day cold gas content
of galaxies?  Suppose that the star formation of a galaxy is simply 
proportional to the mass $\rm M_g$ of cold interstellar matter.  The
rate at which the ISM is consumed is then
$$\rm d M_g ~ = ~ -f M_g ~ dt$$
where f is the fraction of the gas used up per unit time ($\rm
d M_g$ is about 0.7 - 0.8 times the total star formation rate,
because a fraction of the gas is returned by stellar evolution).
If the initial mass of the galaxy is $\rm M_o$, then today ($\rm
t_o$)
$$\rm {{M_g}\over{M_o}} ~ = ~ {{M_g}\over{M_g + M_{\star}}} ~
= ~ e^{-f t_o}$$

The observations discussed above show that elliptical galaxies typically
have $\rm M_g/L_B ~ = ~ 0.002 ~ \rightarrow 0.2 ~ M_{\odot}/L_{\odot}$.
Taking a typical stellar mass to blue light ratio for these galaxies of 
3 $\rm M_{\odot}/L_{\odot}$ gives a median present day value of 
$\rm M_g/M_o$ of about 0.004 for an $L_{\star}$ galaxy, corresponding to
f = $\rm 10^{-9} ~ yr^{-1}$.  The same calculation for a spiral galaxy
gives f = $\rm 2 \times 10^{-10} ~ yr^{-1}$.  These numbers correspond
to a present star formation rate of 1 - 3 $\rm M_{\odot} ~ yr^{-1}$
in spiral galaxies and $\rm M_g/M_o$ = 0.5 at z = 2-- 3 for an $\rm L_{\star}$
elliptical galaxy.

Elliptical galaxies are similar to spiral galaxies, in other words, in
that they seem to have an evolving interstellar medium; the difference is
that the timescale for star formation is much shorter.  Franceschini et al.
(1998) find indications that the star formation timescale is shorter for
galaxies of larger mass.  Could the `second parameter' which determines
the star formation rate be the velocity dispersion of the galaxy or
galaxy subcomponent?

Thus the origin of the gas in elliptical galaxies may be no different
from that in spiral galaxies, i.e. a combination  of intrinsic gas, stellar
mass loss and infall.  This seems not to be the case for elliptical galaxies 
in clusters; these galaxies contain a lot of hot gas but are almost
devoid of cold gas.  The fraction of `peculiar' elliptical galaxies
in the Revised Shapley-Ames Catalogue (Sandage and Tammann 1981)
goes from about 5\% in clusters to almost 50\% in the field.  Perhaps
the epithet `peculiar' has been attached to the wrong galaxies; it looks
as though galaxies in the field, both spirals and ellipticals, usually
contain some coeval cold gas but that galaxies in clusters, both
ellipticals and (former) spirals, do not.
Elliptical galaxies in clusters may be no more typical than are spiral
galaxies in clusters.

\acknowledgments

Patricia Carral, Jordi Cepa and their colleagues at the University of
Guanajuato put together a most educational and enjoyable meeting.  I
thank them for their invitation, and for their most generous support.  
I am greatly indebted to Dave Schiminovich, Sangeeta Malhotra, 
Jacqueline van Gorkom, Michael Rupen and Pat Udomprasert, who
generously allowed me to use thier new results before publication.  I also 
thank Princeton University and NASA, via grant NAG5-3364, for support of this
work, and NASA for its support of the NASA Extragalactic Data Base (NED).

\begin{question}{E. Brinks}
I like your suggestion that one should consider the peculiar Es in the
field as being actually the normal E population and that the Es in clusters
have been stripped of their ISM.  But as Es have quite a deep potential 
well, how can you strip them, and in such an efficient way?
\end{question}
\begin{answer}{G. Knapp}
Perhaps its better to say that the cold gas in elliptical galaxies in
clusters is heated by the same processes as heat the intergalactic gas.
\end{answer}

\begin{question}{R. Bower}
The fraction of the baryons which are in stars in clusters of galaxies
is only about 10\%.  So I agree with the suggestion by Cen and Ostriker
that a similar fraction holds for the field galaxies.
\end{question}
\begin{question} {C. Baugh}
The Cen \& Ostriker argument inferring the global density of stars at
z = 3 from observations of Lyman-break galaxies is weak - the Lyman-break
galaxies sample only a small range of the full luminosity function
at this redshift - a lot more star formation could be going on in smaller
systems.  Also, corrections to the star formation rate for obscuration
by dust could be as much as a factor of 2-5.
\end{question}

\end{document}